\documentclass{elsart}               % The use of LaTeX2e is preferred.

\usepackage{amssymb}                                     % printed Automatica style.

\usepackage{graphicx}          % Include this line if your
\journal{CEJP}

\begin{document}

\begin{frontmatter}

\title{Coordination Control of Heterogeneous Compounded-Order Multi-Agent Systems with Communication Delays\thanksref{footnoteinfo}} % Title, preferably not more

\thanks[footnoteinfo]{This paper is not under consideration for publication with other journals. Corresponding author H.Y. Yang, e-mail: hyyang$\_$ld@yahoo.com.cn.}

\author{Hong-yong Yang$^1$, Lei Guo$^1$, Xun-lin Zhu$^2$, Ke-cai Cao$^3$}

\address{1. National Key Laboratory on Aircraft Control Technology, Beihang University, Beijing 100191, P.R. China; e-mail: $\{$hyyang$\_$ld, lguo66$\}$@yahoo.com.cn.\\
2. School of Mathematics, Zhengzhou University, Zhengzhou, Henan 450001, China; e-mail: hntjxx@163.com.\\
3. School of Automation, Nanjing University of Posts and Telecommunications, Nanjing 210046, China; e-mail: caokc@njupt.edu.cn.
}
%\author[Paestum]{Marcus Tullius Cicero}\ead{cicero@senate.ir},    % Add the
%\author[Rome]{Julius Caesar}\ead{julius@caesar.ir},               % e-mail address
%\author[Baiae]{Publius Maro Vergilius}\ead{vergilius@culture.ir}  % (ead) as shown

%\address[Paestum]{Buckingham Palace, Paestum}  % Please supply
%\address[Rome]{Senate House, Rome}             % full addresses
%\address[Baiae]{The White House, Baiae}        % here.

                             % keyword list or with the
                                          % help of the Automatica
                                          % keyword wizard

\begin{abstract}                          % Abstract of not more than 200 words.
Since the complexity of the practical environment, many distributed networked systems can not be illustrated with the integer-order dynamics and only be described as the fractional-order dynamics. Suppose multi-agent systems will show the individual diversity with difference agents, where the heterogeneous (integer-order and fractional-order) dynamics are used to illustrate the agent systems and compose integer-fractional compounded-order systems. Applying Laplace transform and frequency domain theory of the fractional-order operator, consensus of delayed multi-agent systems with directed weighted topologies is studied. Since integer-order model is the special case of fractional-order model, the results in this paper can be extend to the systems with integer-order models. Finally, numerical examples are used to verify our results.
\end{abstract}

\end{frontmatter}

\section{Introduction}
Nowadays, the past three decades have witnessed significant progress on fractional calculus, because the applications of fractional calculus are found in more and more scientific fields, covering mechanics, physics, engineering, informatics, and materials. The list of such applications is long, for instance, it includes viscoelasticity, colored noise, dielectric polarization, electrode-electrolyte polarization, electromagnetic waves, control engineering and so on\cite{podlubny,hilfer,ladaci,hwanga,ahn,li,sabat,lu}. In fact, real word processes generally or most likely are fractional-order systems\cite{podlubny,hilfer}. Furthermore, fractional order controllers have so far been implemented to enhance the robustness and the performance of the closed loop control systems, the stability problem of fractional-order systems has been investigated both from an algebraic and an analytic point of view\cite{ladaci,hwanga,ahn,li,sabat,lu}.

On the other hand, as one of the most basic problem of coordination control for networked systems, consensus of multi-agent systems has been widespread concerned as an important research topic in the field of systems control. Consensus of multi-agent systems means that several distributed agents achieve the same state or output through local mutual coupling effect among the individuals, where the centralized control is not used. Based on the computer model proposed by Reynolds which imitates animals' flocking \cite{5}, Vicsek et al. \cite{6} firstly proposed a non-equilibrium multi-agent system model from the point of view of statistical mechanics, simulation shows that all individuals in the system can run in accordance with the same direction under certain conditions. Recently, since the wide application of multi-agent systems in various fields, many scholars have devoted themselves to study the consensus of multi-agent systems \cite{olfati,ren07,lishihua,lin,yujun,yang,tian}. When the information transferring, communication delays will occur in networked control system and affect the features of the system. The effects of communication delays on movement consensus of multi-agent systems have been concerned by many scholars \cite{lin,yujun,yang,tian}, and the stability of delay system also become a hot topic in the multiple agents field.

The important results of the above literatures pay attention to the consensus problem of integer-order multi-agent systems. In the complex environment, many dynamic characteristics of natural phenomena can not be described in the form integer-order equation, but only be described in the dynamics of fractional-order (non-integer order) behavior, for example: flocking movement and food searching by means of the individual secretions and microbial, submarine underwater robots in the bottom of the sea with a large number of microorganisms and viscous substances, unmanned aerial vehicles running in the complex space environment \cite{ren2011}. Cao and Ren  \cite{cao1,cao2} studied distribution coordination of fractional-order multi-agent systems firstly, and gave the relationship between the number of individuals and the fractional order in the stable multi-agent systems. However, to the best of authors' knowledge, there are few researches done on the coordination control of fractional-order multi-agent systems with communication delays.

In this paper, we suppose agents work in the complex environment, the heterogeneous dynamics with fractional-order and integer-order is presented for multi-agent systems. The main innovation of this paper lies in the study on consensus of compound-order (fractional-order and integer-order) distributed multi-agent systems with different time delays. This paper is organized as follows. In Section 2, some necessary definitions and notations are given on fractional calculus. In Section 3, the compound-order dynamics of the fractional order system and the integer order system is presented. The consensus of integer-fractional compounded-order multi-agent systems with communication delays is studied in Section 4. The corresponding simulation results are provided in Section 5 to demonstrate the effectiveness of the proposed conditions. Finally, the conclusions are drawn in Section 6.

\section{Fractional Calculus}
Fractional calculus plays an important role in modern science. There are mainly two widely used fractional operators: Caputo and Riemann-Liouville (R-L) fractional operators\cite{podlubny}. In physical systems, Caputo fractional operator is more practical than R-L fractional operator because R-L fractional operator has initial value problems. Therefore, in this paper we will use Caputo fractional operator to model the system dynamics and analyze the stability of the proposed coordination algorithms. Generally, Caputo fractional operator includes Caputo integral and Caputo derivative. Caputo integral is defined as
\begin{eqnarray*}
^{C}_{a}D^{-p}_{t}f(t)=\frac{1}{\Gamma(p)}\int^{t}_{a}\frac{f(\theta)}{(t-\theta)^{1-p}}d \theta,
\end{eqnarray*}
where the integral order  $p\in (0,1]$, $\Gamma(.)$ is the Gamma function, and a is an arbitrary real number. Based on the Caputo integral, for a nonnegative real number $\alpha$, Caputo derivative is defined as
\begin{equation}
^{C}_{a}D^{\alpha}_{t}f(t)=^{C}_{a}D^{-p}_{t}[\frac{d^{[\alpha]+1}}{dt^{[\alpha]+1}}f(t)],
\label{e1}
\end{equation}
where $p=[\alpha]+1-\alpha\in (0,1]$ and $[\alpha]$ is the integral part of $\alpha$. If $\alpha$ is an integer, then $p=1$ and the Caputo derivative is equivalent to the integer-order derivative. In this paper, a simple notation $f^{(\alpha)}(t)$ is used to replace $^{C}_{a}D^{\alpha}_{t}f(t)$. Let $\mathfrak{L}()$ denote the Laplace transform of a function, the Laplace transform of Caputo derivative is shown as
\begin{equation}
\mathfrak{L}(f^{(\alpha)}(t))=s^{\alpha}F(s)-\sum^{[\alpha]+1}_{k=1}s^{\alpha-1}f^{(k-1)}(0),
\label{e2}
\end{equation}
where $F(s)=\mathfrak{L}(f(t))=\int^\infty_{0-}e^{-st}f(t)dt$ is the Laplace transform of function $f(t)$, $f^{(k)}(0)=\lim_{\xi\rightarrow 0-}f^{(k)}(\xi)$ and $f^{(0)}(0)=f(0)=\lim_{\xi\rightarrow 0-}f(\xi)$.

%\begin{figure}
%\begin{center}
%\includegraphics[height=4cm]{jcaesar.eps}    % The printed column
%\caption{Gaius Julius Caesar, 100--44 B.C.}  % width is 8.4 cm.
%\label{fig1}                                 % Size the figures
%\end{center}                                 % accordingly.
%\end{figure}

% OR

%\begin{figure}
%\begin{center}
%\epsfig{file=jcaesar,width=7cm}
%\caption{Gaius Julius Caesar, 100--44 B.C.}
%\label{fig1}
%\end{center}
%\end{figure}

\section{Problem statement}

Assume that multi-agent systems consist of $n$ autonomous agents, connected relations among the agents constitute a network topology $\mathcal{G}$. Let $\mathcal{G}=\{V,E,A\}$ represent a directed weighted graph, in which $V=\{v_1,v_2,...,v_n\}$ represents a collection of $n$ nodes, and its set of edges is $E\subseteq V\times V$. The node indexes belong to a finite index set
$I=\{1,2,...,n\}$, adjacency matrix $A=[a_{ij}]\in R^{n\times n}$ with weighted adjacency elements
$a_{ij}\geq 0$. An edge of the weighted diagraph $\mathcal{G}$ is
denoted by $e_{ij}=(v_{i},v_{j})\in E$. We assume that the adjacency
element $a_{ij}> 0$ when $e_{ij}\in E$, otherwise, $a_{ij}= 0$. The
set of neighbors of node $i$ is denoted by $N_{i}=\{j\in I:
a_{ij}>0\}$.

Let $\mathcal{G}$ be a weighted digraph without self-loops, i.e.,
$a_{ii}=0$, and matrix $D=\mathrm{diag}\{d_1,d_2,...,d_n\}$ be the diagonal matrix with the diagonal
elements $d_{i}=\sum_{j=1}^{n}a_{ij}$ representing the sum of the elements in the $i$-th row of matrix $A$.
The Laplacian matrix of the weighted digraph $\mathcal{G}$ is defined as
$L=D-A$. For two nodes $i$ and $k$, there is subscript set $\{k_{1},
k_{2}, ... k_{l}\}$ satisfying $a_{ik_{1}}>0$, $a_{k_{1}k_{2}}>0$,
..., $a_{k_{l}k}>0$, then there is a directed linked path between node $i$ and node $k$ which is used for the information transmission, also we can say node $i$ can receive the information from node $k$. If node $i$ can find a path to reach any node of the graph, then node $i$ is globally reachable from every other node in the digraph. For any two nodes in the graph, there are at least one directed linked path, then G is strongly connected.

{\bf Lemma 1}\cite{ren07}. $0$ is a simple eigenvalue of Laplacian matrix $L$, and
$X_{0}=C[1, 1, ..., 1]^{T}$ is corresponding right eigenvector,
i.e., $LX_{0}=0$, if and only if the digraph $\mathcal{G}=(V, E,
A)$ has a globally reachable node.

Assume that there are individual differences in the complex environment of multi-agent systems; there are two groups of  multi-agent systems with integer-order dynamics and fractional-order dynamics. The compounded-order dynamical equations are described as follows:

\begin{equation}
\begin{array}{lc}
\dot{x}_{i}(t)=u_i(t), i=1,...,m,\\
x_{l}^{(\alpha)}(t)=u_l(t), l=m+1,...,n,
\end{array}\label{e3}
\end{equation}
where $x_i(t)\in R$ and $u_i(t)\in R$ represent the $i$-th agent's state and control input respectively, $\dot{x}_{i}(t)$ represents the first-order derivative for the state $x_i(t)$, $x_{l}^{(\alpha)}$ represents the $\alpha $ order Caputo derivative, and $\alpha\in (0, 1]$. Assume the following control protocols are used in multi-agent systems:
\begin{equation}
u_{i}(t)=-\gamma\sum_{k\in N_i}a_{ik}[x_{i}(t)-x_{k}(t)], i=1,...,m,m+1,...,n.
\label{e4}
\end{equation}
where $a_{ik}$ represents the $(i, k)$ elements of adjacency matrix $A$, $\gamma>0$ is control gain, $N_i$ represents the neighbors collection of the $i$-th agent.

This article assumes that there are communication delays in the dynamical systems, and consensus of the integer-fractional-compounded-order agent systems with communication delays is studied. Under the influence of communication delays, we can get the following algorithm:
\begin{equation}
\begin{array}{lc}
\dot{x}_{i}(t)=u_i(t-\tau_i), i=1,...,m,\\
x_{l}^{(\alpha)}(t)=u_l(t-\tau_l), l=m+1,...,n,
\end{array}
\label{e5}
\end{equation}
where $\tau_i>0$ is the communication delay of agent $i$. Through a simple change we can get
\begin{equation}
[\dot{X}_1(t),X_2^{(\alpha)}]^T=-\gamma[\Sigma_{i=1}^{m}L_{i}X^T(t-\tau_i), \Sigma_{i=m+1}^{n}L_{i}X^T(t-\tau_i)]^T
\label{e6}
\end{equation}
where $L_i=E(i)L$, $E(i)$ represents matrix whose element of $(i, i)$ is 1 and the rest are 0s, $X_1(t)=[x_1(t),...,x_m(t)]$, $X_2(t)=[x_{m+1}(t),...,x_n(t)]$, $X(t)=[x_1(t),x_2(t),...,x_n(t)]$, and $L=\Sigma_{i=1}^{n}L_{i}$. Suppose that for any initial value of the system, the states of autonomous agents meet $\lim_{t\rightarrow\infty}(x_i(t)-x_k(t))=0$, for $i,k\in I$, then we call multi-agent systems asymptotically reach consensus. In this paper, we apply Laplace transform and frequency domain theory of the fractional-order operator to study the consensus of delayed compounded-order multi-agent systems with directed weighted topologies.

\section{Coordination control for compounded-order multi-agent systems with communication-delays}
%\subsection{Disturbance observer}
{\bf Theorem 1} Suppose that multi-agent systems are composed of $n$ independent agents whose connected network topology is directed and has a globally reachable node. Then, compounded-order multi-agent system (\ref{e6}) with time delays can asymptotically reach consensus, if
\begin{equation}
\tau_i<\frac{\pi}{2(2\gamma \bar{d})^{1/\alpha}}
\label{e7}
\end{equation}
where $\bar{d}=\max\{d_i, i=1,...,n\}$, $d_i=\Sigma_{k=1}^{n}a_{ik}$.

{\bf Proof.} Applying Laplace transformation to system (\ref{e6}), the characteristic equation can be gotten
\begin{equation}
\det(\left(\begin{array}{cc}sI_m\ \ &  \\  & s^\alpha I_{n-m} \end{array}\right)+\gamma E(s)L)=0,
\label{e8}
\end{equation}
where $I_m$ represents a unit matrix with $m$-dimensions, $E(s)=\mathrm{diag}\{e^{-\tau_1 s},...,e^{-\tau_n s}\}$. 0 is a single eigenvalue of the Laplacian matrix $L$ because the system has a globally reachable node from Lemma 1. Due to $\alpha>0$, the characteristic equation has a characteristic root $s=0$.

When $s\neq 0$, let
\begin{eqnarray*}F(s)=(I_n+\gamma \left(\begin{array}{cc}s^{-1}I_m\ \ &  \\  & s^{-\alpha} I_{n-m} \end{array}\right) E(s)L),\end{eqnarray*}
the characteristic equation (\ref{e8}) is equivalent to $F(s)=0$. Nextly, we will prove that all zero points of $F(s)=0$ have negative real parts. Let \begin{eqnarray*}G(s)=\gamma \left(\begin{array}{cc}s^{-1}I_m\ \ &  \\  & s^{-\alpha} I_{n-m} \end{array}\right)E(s)L,\end{eqnarray*} according to the generalized Nyquist criterion \cite{desoer}, if for $s=j\omega$, where $j$ is complex number unit, point $-1+j0$ is not surrounded by the Nyquist curve of $G(j\omega)$'s eigenvalues, then all zero points of $F(s)$ have negative real parts. Let $s=j\omega$, we can get
\begin{equation}
G(j\omega)=\gamma \left(\begin{array}{cc}\omega^{-1} e^{-j\pi/2}I_m\ \ &  \\  & \omega^{-\alpha}e^{-j\alpha\pi/2}I_{n-m} \end{array}\right)\ E(j\omega)L.
\label{e9}
\end{equation}
In the following proof, Gerschgorin's disc theorem will be applied to estimate the eigenvalues $\lambda(G(j\omega))$ of the matrix $G(j\omega)$. According to the Gerschgorin's disc theorem, we have
\begin{equation}
\lambda(G(j\omega))\in \bigcup_{i\in I} G_i,
\label{e10}
\end{equation}

where for $i=1,...,m$,
\begin{eqnarray*}
\begin{array}{cl}
G_i=& \{\zeta\in C, |\zeta-\omega^{-1}\gamma d_{i}e^{-j(\omega\tau_i+\pi/2)}|\\
& \leq \omega^{-1}\sum_{k=1,k\neq i}^{n}|\gamma a_{ik}e^{-j(\omega\tau_i+\pi/2)}|\},
\end{array}
\end{eqnarray*}
and for $i=m+1,...,n$,
\begin{eqnarray*}
\begin{array}{cl}
G_i=& \{\zeta\in C, |\zeta-\omega^{-\alpha}\gamma d_{i}e^{-j(\omega\tau_i+\alpha\pi/2)}|\\
& \leq \omega^{-\alpha}\sum_{k=1,k\neq i}^{n}|\gamma a_{ik}e^{-j(\omega\tau_i+\alpha\pi/2)}|\},
\end{array}
\end{eqnarray*}
where $d_{i}=\sum_{k=1}^{n}a_{ik}$. After simple sorting it can be
\begin{eqnarray*}
\begin{array}{cl}
G_i=& \{\zeta\in C, |\zeta-\omega^{-1}\gamma d_{i}e^{-j(\omega\tau_i+\pi/2)}|\leq \omega^{-1}\gamma d_{i}\}, i=1,...,m,
\end{array}
\end{eqnarray*}
and
\begin{eqnarray*}
\begin{array}{cl}
G_i=& \{\zeta\in C, |\zeta-\omega^{-\alpha}\gamma d_{i}e^{-j(\omega\tau_i+\alpha\pi/2)}|\leq \omega^{-\alpha} \gamma d_{i}\}, i=m+1,...,n,
\end{array}
\end{eqnarray*}

When the Nyquist curve of the origin of the disc $G_i$ changing, the disc changes along with it. Next, we will prove that point $-a+j0 (a\geq 1)$ is not in every disc $G_i $.

For $i=1,...,m$, the changes of the disc $G_i$ is to be studied in the following. The origin of the disc $G_i$ is $\omega^{-1}\gamma d_{i}e^{-j(\omega\tau_i+\pi/2)}$, the radius of the disc $G_i$ is $\omega^{-1}\gamma d_{i}$. Let
\begin{eqnarray*}
\Delta=|-a+j0-\omega^{-1}\gamma d_{i}e^{-j(\omega\tau_i+\pi/2)}|^2-(\omega^{-1}\gamma d_{i})^2,
\end{eqnarray*}
there is
\begin{eqnarray*}
\Delta=a(a+2\omega^{-1}\gamma d_{i}\cos(\omega\tau_i+\pi/2)).
\end{eqnarray*}
When $\omega_c \tau_i+\pi/2=\pi$, there is $\cos(\omega_c\tau_i+\pi/2)=-1$. We can get
\begin{eqnarray*}
\Delta\geq a(a-2\omega_c^{-1}\gamma d_{i}),
\end{eqnarray*}
where
\begin{eqnarray*}
2\omega_c^{-1}\gamma d_{i}=2(\pi/(2\tau_i))^{-1}\gamma d_i.
\end{eqnarray*}
According to the conditions of the theorem
\begin{eqnarray*}
2(\pi/(2\tau_i))^{-1}\gamma d_i\leq 2((\pi/(2\tau_i))^{-\alpha}\gamma \bar{d}<1,
\end{eqnarray*}
from the hypothesis $a\geq 1$, we can get
\begin{eqnarray*}
\Delta>0.
\end{eqnarray*}
Then, it gets, for $i=1, ..., m$,
\begin{eqnarray*}
|-a+j0-\omega^{-1}\gamma d_{i}e^{-j(\omega\tau_i+\pi/2)}|>\omega^{-1}\gamma d_{i}.
\end{eqnarray*}

By means of the same deduction, we can get, for $i=m+1, ..., n$,
\begin{eqnarray*}
|-a+j0-\omega^{-\alpha}\gamma d_{i}e^{-j(\omega\tau_i+\alpha\pi/2)}|>\omega^{-\alpha}\gamma d_{i}.
\end{eqnarray*}

When $a\geq 1$, the point $-a+j0$ is not in disc $G_i$. Thus the point $-1+j0$ is not surrounded by curves of eigenvalue $\lambda(G(j\omega))$ of matrix $G(j\omega)$. Therefore, all zero points of $F(s)=0$ have negative real parts.

Due to the equilibrium point of the system meeting $LX^*=0$, then $X^*=C[1,...,1]^T$ (where $C$ is a constant) is the eigenvector with the corresponding eigenvalue 0 of the Laplacian matrix $L$. Therefore, $\lim_{t\rightarrow\infty}x_{i}(t)=C$, and the system asymptotically reaches consensus.

{\bf Corollary 1.} Suppose multi-agent systems are composed of $n$ independent agents, whose connection network topology is directed and symmetrical, and there is a global reachable node. Then compounded-order multi-agent system (\ref{e6}) with time delays can asymptotically reach consensus, if
\begin{equation}
\tau_i<\frac{\pi}{2(\gamma \rho_{L})^{1/\alpha}},
\label{e11}
\end{equation}
where $\rho_L$ represents the spectral radius of matrix $L$ with $\rho_L=\max\{|\lambda_i|,i=1,...,n\}$ and $\lambda_i$ is the eigenvalue of the Laplacian matrix $L$.

{\bf Proof.} According to theorem 1, the characteristic equation of the system is
\begin{eqnarray*}
\det(\left(\begin{array}{cc}sI_m &  \\  &  s^\alpha I_{n-m} \end{array}\right)+\gamma E(s)L)=0.
\end{eqnarray*}
Since the Laplacian matrix $L$ is symmetrical, there is orthogonal matrix $P$ satisfying $L=P\Lambda P^{-1}$, where $\Lambda=diag\{ \lambda_1,...,\lambda_n\}$. Because there is a global reachable point, we can know $Rank(L)=n-1$ and 0 is a single eigenvalue of matrix   from Lemma 1. Therefore, the characteristic equation has a root $s=0$.

When $s\neq 0$, let $F(s)$ and $G(s)$ be same as the proof of Theorem 1, and let
\begin{equation}
H(s)=\left(\begin{array}{cc}s^{-1}I_m &  \\  &  s^{-\alpha} I_{n-m} \end{array}\right)\mathrm{diag}\{e^{-\tau_is}, i=1,...,n\}.
\label{e12}\end{equation}

Let $s=j\omega$, we can get
\begin{equation}
G(j\omega)=H(j\omega)\gamma L,
\label{e13}
\end{equation}
where
\begin{eqnarray*}
\begin{array}{cl}
H(j\omega)&=\mathrm{diag}\{H_i(j\omega),i=1,...,n\}\\
&=\left(\begin{array}{cc}\omega^{-1}e^{-j\pi/2} &  \\  &  \omega^{-\alpha}e^{-j\alpha\pi/2} I_{n-m} \end{array}\right)\mathrm{diag}\{e^{-j\omega\tau_i}, i=1,...,n\}.

\end{array}\end{eqnarray*}

Let $M=\mathrm{diag}\{M_i, i=1,...n\}$ where $M_i=\pi/(2\tau_i)$ (for $i=1,...,m$) and $M_l=((2-\alpha)\pi/(2\tau_l))^\alpha$ (for $l=m+1,...,n$).
Matrix $MH(j\omega)=\mathrm{diag}\{M_iH_i(j\omega),\ i=1,...,n\}$ is a diagonal matrix where the Nyquist curve of its diagonal elements passes over point $-1+j0$. Suppose $\lambda(G(j\omega))$ is the eigenvalue of matrix $G(j\omega)$,
we have
\begin{eqnarray*}
\begin{array}{cl}
\lambda(G(j\omega))& =\lambda(MH(j\omega)\gamma M^{-1/2}LM^{1/2})\\
& \in \rho(\gamma M^{-1/2}LM^{1/2}) \times Co(0\cup \{M_iH_i(j\omega),i=1,...,n\}),
\end{array}
\end{eqnarray*}
Where $\rho()$ represents the spectral radius of matrix, $Co(\xi)$ represents the convex hull of $\xi$. Because of $M_iH_i(j\omega)$ will passes over point $-1+j0$, point $-1+j0$ is included in convex hull $Co(0\cup \{M_iH_i(j\omega), i=1,...,n\})$. Since
\begin{eqnarray*}
\begin{array}{cl}
\rho(\gamma M^{-1/2}LM^{1/2})&= \rho(\gamma M^{-1/2}\Lambda M^{1/2})\\
& =\max\{|\gamma M_i^{-1}\lambda_i|,i=1,...,n\},
\end{array}
\end{eqnarray*}
according to the hypothesis condition $\tau_i<\frac{\pi}{2(\gamma \rho_{L})^{1/\alpha}}$, we can get
\begin{eqnarray*}
\begin{array}{cl}
\rho(\gamma M^{-1/2}LM^{1/2})&<1.
\end{array}
\end{eqnarray*}
Therefore, point $-1+j0$ is not included in $\rho(\gamma M^{-1/2}LM^{1/2}) \times Co(0\cup \{M_iH_i(j\omega),i=1,...,n\})$. That is, point $-1+j0$  is not included in the Nyquist curve of the eigenvalue of $G(j\omega)$. According to generalized Nyquist theorem \cite{desoer}, the zero points of the characteristic equation have negative real parts. Therefore the multi-agent systems can asymptotically reach consensus, and $\lim_{t\rightarrow\infty}x_{i}(t)=C$.

{\bf Corollary 2.}  Suppose multi-agent systems are composed of $n$ independent agents, whose connection network topology is directed and symmetrical, and there is a global reachable node. Then compounded-order multi-agent system (\ref{e6}) with time delays can asymptotically reach consensus with $\alpha=1$, if
\begin{equation}
\gamma\tau_i<\pi/(2\lambda_{max}),
\label{e14}
\end{equation}
where $\lambda_{max}$ is the maximum eigenvalue of matrix $L$.

{\bf Corollary 3.} Suppose multi-agents are system composed of $n$ independent agents, whose connection network topology is directed and symmetrical, and there is a global reachable node. Then compounded-order multi-agent system (\ref{e6}) can asymptotically reach consensus when $\alpha=1$ and time delays $\tau_i=\tau$, if
\begin{equation}
2\gamma\tau<\pi/\rho,
\label{e15}
\end{equation}
where $\rho$ represents Spectral radius of matrix $L$.

{\bf Remark 1.} The consensus result in Corollary 3 for $\gamma=1$ is in accord with that in \cite{olfati}.

\section{Examples Simulations}
Suppose the system is composed of four dynamical agents (Fig. 1) with two integer-order agent systems (agnet1 and agent2) and two fractional-order agent systems (agent3 and agent4). The connection weights between individuals are $a_{21} = 0.7$, $a_{42} = 0.8$, $a_{31} = 0.9$, $a_{14} = 1$. The order of the fractional multi-agent dynamics is $\alpha=0.9$, through the topology of the system,we can get the adjacency matrix
\begin{eqnarray*}
A=\left( \begin{array}{cccc}
0&  0&   0&  1\\
0.7 & 0 & 0 & 0\\
0.9& 0& 0 & 0\\
0& 0.8 & 0& 0
\end{array}
\right).
\end{eqnarray*}

%\begin{figure}\center %\epsfxsize=8cm \epsfysize=6cm
%\includegraphics[height=2in,width=3in]{fig1.eps}
%\begin{center}{Fig.~1:  Network topology of the multi-agent systems. }
%\end{center}   \end{figure}

\begin{center}
 \scalebox{0.5}[0.25]{\includegraphics{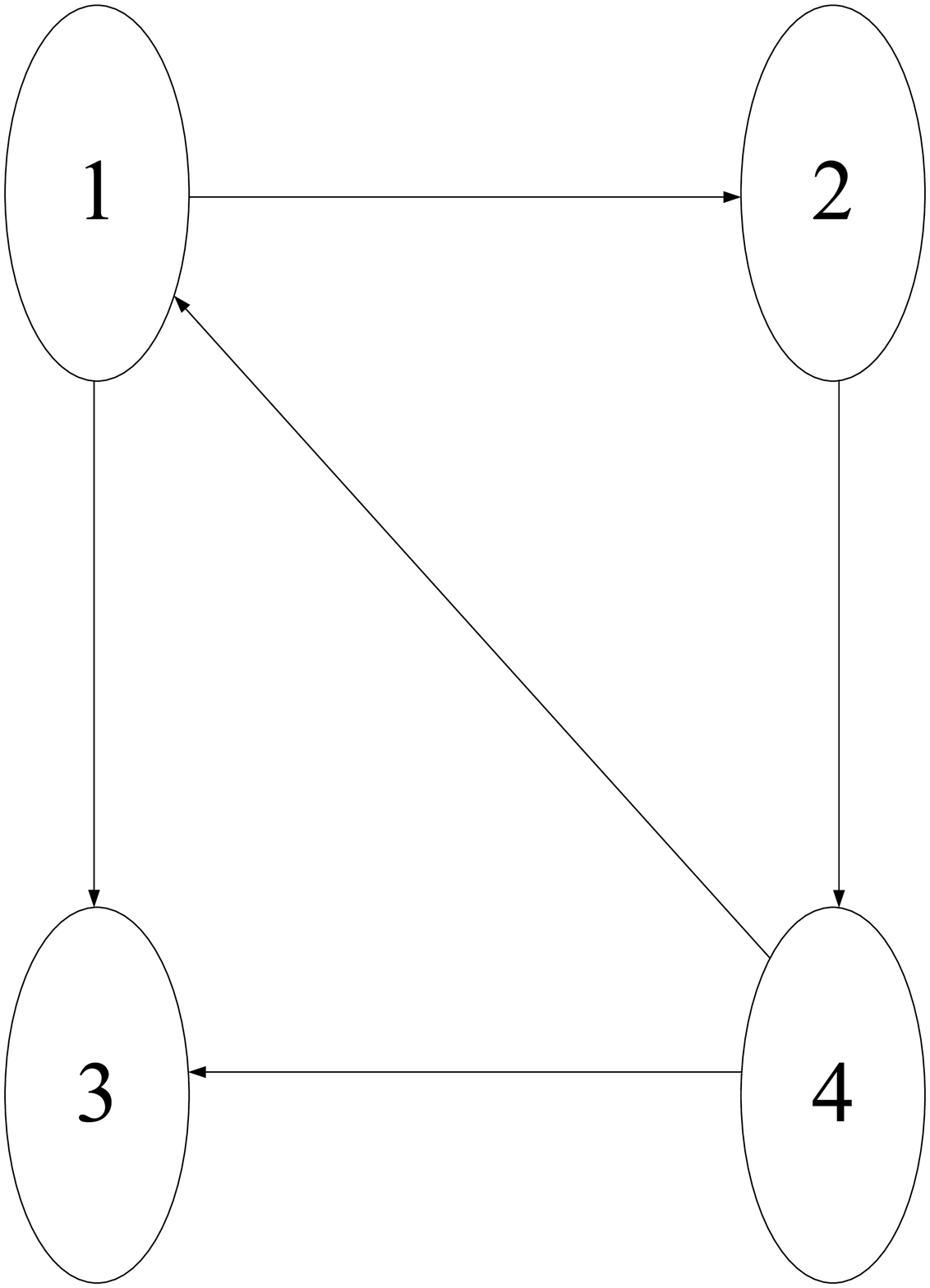}}
\center{\small Fig. 1. Network topology of the multi-agent systems.}
\end{center}

According to Theorem 1, we can get the relationship between the system control gain and the upper bound of communication delays (Fig. 2). With the help of Fig.2, we can select the control gain according to the communication delay of the system, or decide the upper bound of communication delays by means of the control gain of the system, to make the system meet the condition of reaching consensus. Suppose the communication delay is $\tau=0.6s$, the system control gain should be selected as $\gamma \leq 1.19$ from Fig. 2; suppose the system control gain $\gamma=1$, we can obtain that the upper bound of communication delays is $0.7s$ from Fig.2.

Assume the communication delay of multi-agent systems is 0.6s and the system control gain $\gamma=1$ in simulation, we set the expect objective at 0.5, consensus can be asymptotically reached (Fig. 3) through compounded-order coordination algorithm.

\begin{figure}\center %\epsfxsize=8cm \epsfysize=6cm
\includegraphics[height=3in,width=4in]{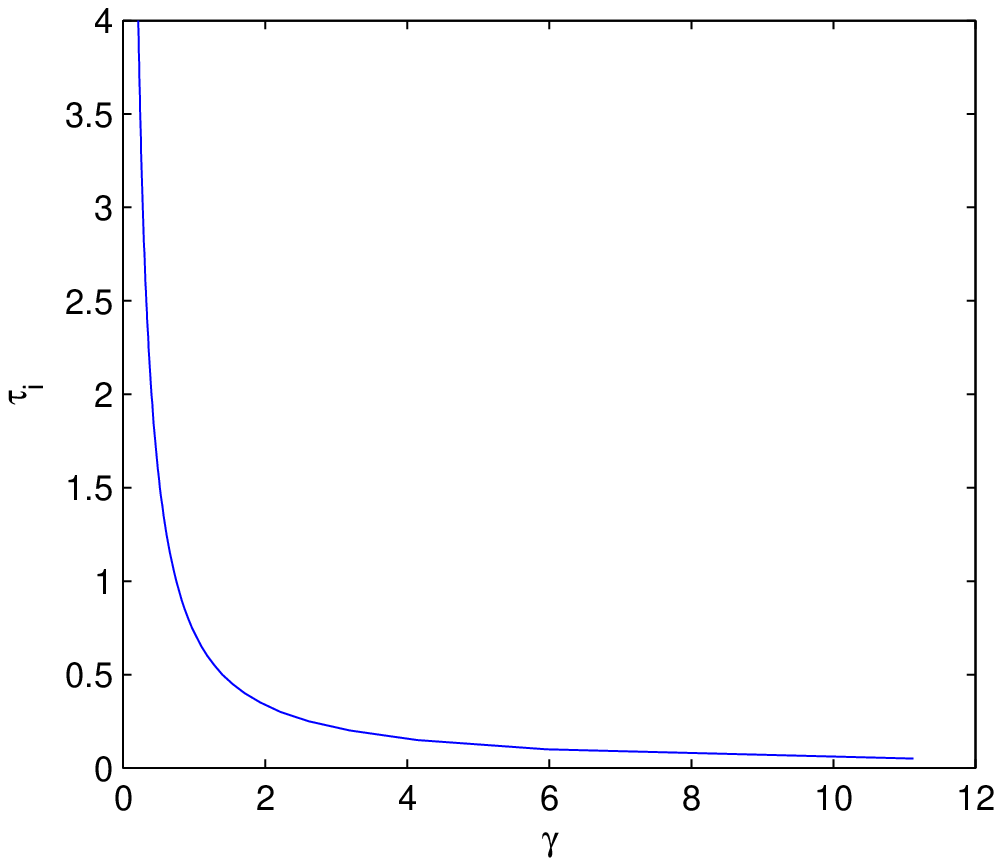}
\begin{center}{Fig.~2:  Relationship between the control gain and the upper bound of communication delays. }
\end{center}   \end{figure}

\begin{figure}\center %\epsfxsize=8cm \epsfysize=6cm
\includegraphics[height=3in,width=4in]{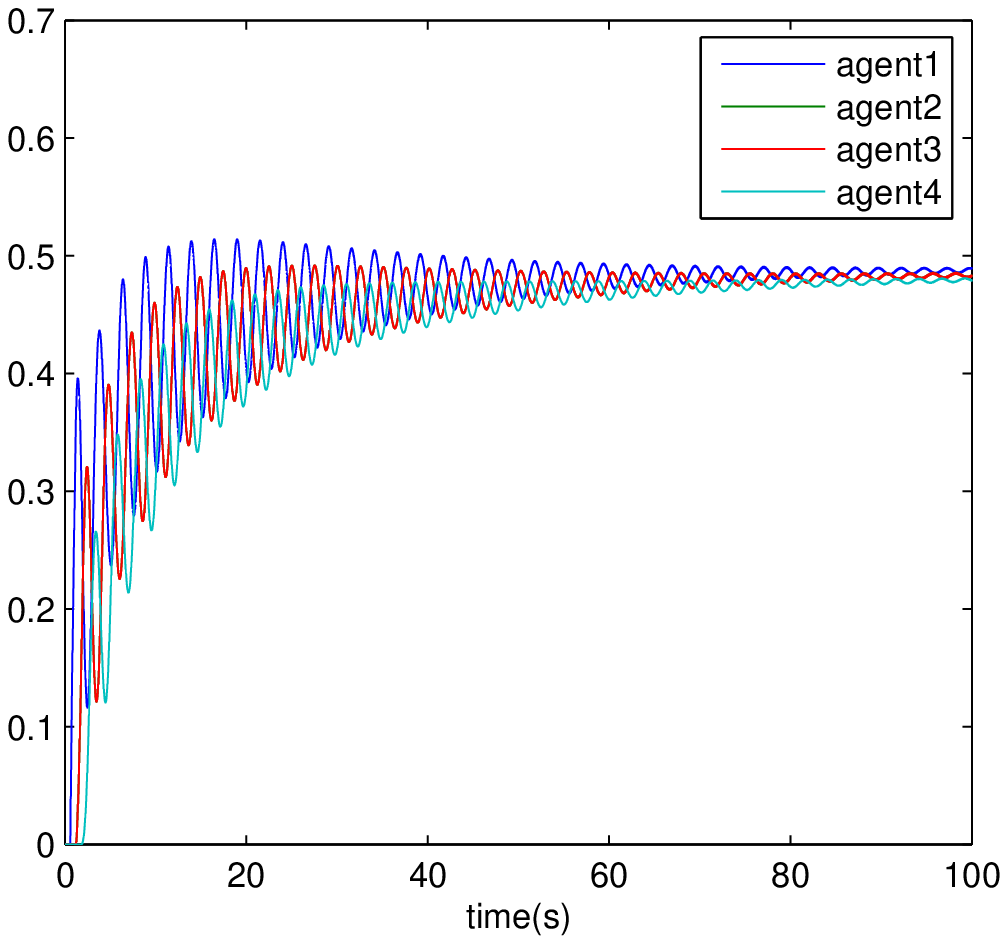}
\begin{center}{Fig.~3:  Movement trajectories of the multi-agent systems with delay 0.6s. }
\end{center}   \end{figure}

Assume the communication delay of multi-agent systems is 0.7s and the system control gain $\gamma=1$ in simulation, consensus of compounded-order multi-agent systems can not be reached (Fig. 4).

\begin{figure}\center %\epsfxsize=8cm \epsfysize=6cm
\includegraphics[height=3in,width=4in]{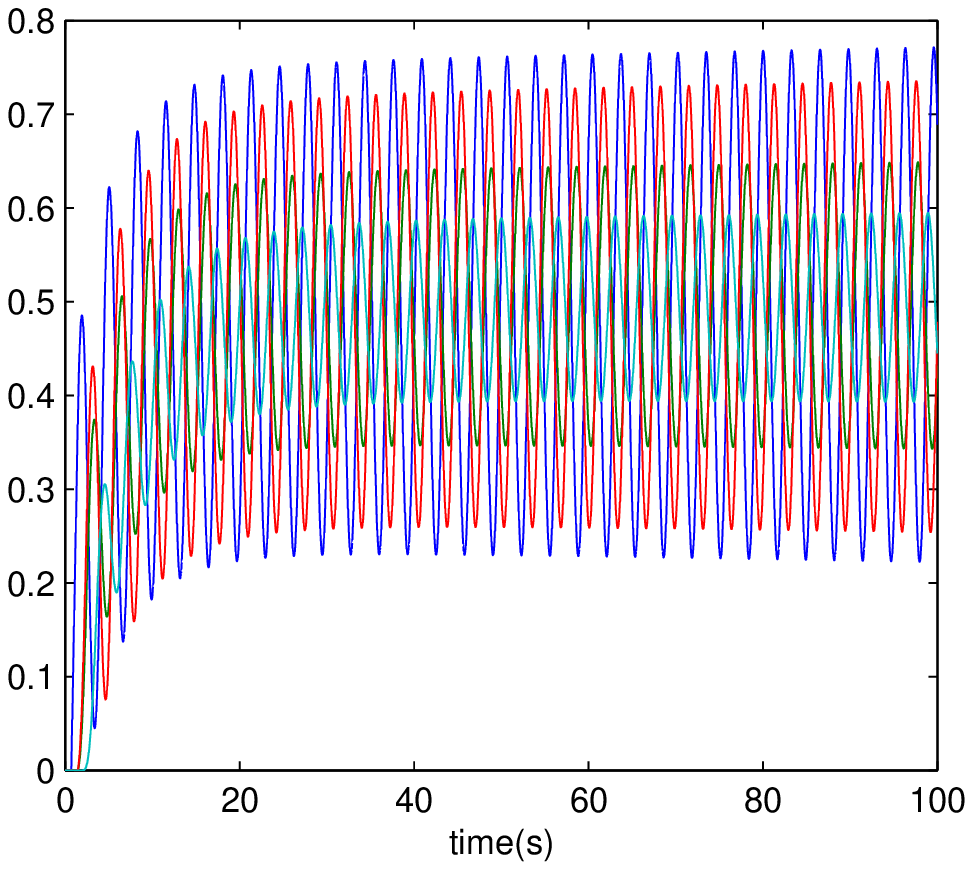}
\begin{center}{Fig.~4:  Movement trajectories of the multi-agent systems with delay 0.7s. }
\end{center}   \end{figure}

\section{Conclusions}
This paper studies distributed coordination of integer-fractional-compounded-order multi-agent systems with communication delays. Consensus of multi-agent systems with directed network topology is studied through the stability theory of frequency domain, and the consensus conditions for compounded-order delayed multi-agent systems are presented. The relationship between the control gain of multi-agent systems and the upper bound of time delays is derived. Suppose the orders of the fractional dynamical systems are all 1, the extended conclusion in this paper is the same with ordinary integer order system. In the following work, research of the robust stability of integer-fractional-compounded-order multi-agent systems will be carried out.

\section*{Acknowledgements}                               % Place acknowledgements
This research is supported in part by the State Key Development Program for Basic Research of China (Grant No. 2012CB720003), the National Natural Science Foundation of China (Grant No. 91016004, 61273152, 61127007) and the Natural Science Foundation of Shandong Province of China (No. ZR2011FM017).

%\begin{ack}                               % Place acknowledgements
%This research is supported by the National Natural Science
%Foundation of China (under grant ), the Science Foundation of Education Office of Shandong
%Province of China (under grant J08LJ01) and Internal Visiting
%\end{ack}

\bibliographystyle{plain}        % Include this if you use bibtex
%\bibliography{autosam}           % and a bib file to produce the
                                 % bibliography (preferred). The
                                 % correct style is generated by
                                 % Elsevier at the time of printing.

\appendix
%\section{A summary of Latin grammar}    % Each appendix must have a short title.
%\section{Some Latin vocabulary}         % Sections and subsections are supported
                                        % in the appendices.
\end{document}